\documentclass[aps,twocolumn,floats,prd,nofootinbib,superscriptaddress,showpacs,10pt]{revtex4-1}

\usepackage{amsfonts,amsmath,amssymb,amsthm}
\usepackage[bookmarks, bookmarksopen]{hyperref}
\usepackage{graphicx}
\usepackage[english]{babel}
\usepackage{physics}

\begin{document}
	\title{Static spherically symmetric configurations with {\it N} non-linear scalar fields: global and asymptotic properties.}

\date{\today}

\author{V.~I.~\surname{Zhdanov} }\thanks{e-mail:  valeryzhdanov@gmail.com}
\affiliation{Taras Shevchenko National University of Kyiv, Kyiv 01033 Ukraine}
\affiliation{Igor Sikorsky Kyiv Polytechnic Institute, Kyiv, 03056 Ukraine}
\author{O.~S.~\surname{Stashko}}\thanks{e-mail: alexander.stashko@gmail.com}
\affiliation{Taras Shevchenko National University of Kyiv, Kyiv 01033 Ukraine}

\begin{abstract}
 In case of a spherically symmetric non-linear scalar field (SF) in flat space, besides singularity at the center, spherical singularities can  occur for non-zero values of radial variable $r>0$. We show that in the General Relativity   the gravitational field suppresses the occurrence of the spherical singularities under some generic conditions. Our consideration deals with  asymptotically flat space-times around static spherically symmetric configurations in presence of $N$ non-linear SFs, which are  minimally coupled to gravity. Constraints are imposed on the  SF potentials, which guarantee a monotonicity of the fields as functions of $r$; also  the potentials are assumed to be  exponentially bounded. We give direct proof  that  solutions of the joint system of Einstein -- SF equations satisfying the conditions of  asymptotic flatness are regular for all values of $r$,  except for naked singularities in the center $r=0$ in the Schwarzschild (curvature) coordinates. Asymptotic relations for SF and metric near the center are  derived, which appear to be remarkably similar to the Fisher solution for free SF. These relations determine two main types of the corresponding geodesic structure when photons can be captured by the singularity or not. To illustrate, the case of one SF with monomial potential is analyzed in detail numerically. We show that the image of the accretion disk around the singularity, observed from infinity, can take the form of a bright ring with a dark spot in the center, like an ordinary black hole.
\end{abstract}


\maketitle

\section{Introduction} \label{Introduction}
In this paper we study some common properties  of static spherically symmetric configurations of General Relativity with  several non-linear scalar fields.

Scalar field (SF) models are widely used in  relativistic gravitational physics as elements of alternative gravitational theories \cite{WillBerti}, especially in cosmology within approaches to the dark energy problem   \cite{DarkEnergy}. {Most publications deal with one SF, however scenarios with multiple SFs are also well known  (see, e.g., \cite{Multifield}), being  extensions of the single-scalar-tensor approaches.}

Therefore, it is natural to ask how the scalar field works in compact astrophysical objects.  In particular, recent publications of the Event Horizon Telescope (EHT) results \cite{EHT1} have  increased  interest in   black hole (BH) mimickers, which differ from BH but  can give similar images of the radiating material around these objects. Indeed, within the EHT angular resolution it is difficult to rule out a number  of  alternative astrophysical objects   (see, e.g., \cite{EHT5Mizuno2018Bambi2019}) and references therein). Theories with additional scalar fields, which model the ubiquitous "Dark Energy", create a suitable soil where such mimickers can grow \cite{Virbhadra_Ellis, Gyulchev2019}.

It is well known that introduction of  SF in models of compact astrophysical objects may lead to important consequences. For example, the arbitrarily small free SF affects the space-time topology, e.g., leading to a naked singularity (NS). This is clearly seen in case of the  Fisher analytic solution \cite{Fisher, Janis,WymanVirbhadra}. This is closely related to the famous Bekenstein theorems  \cite{Bekenstein1972} (see, also \cite{Bekenstein} and references therein), which prohibit existence of horizons in presence of SF.
Though,  space-times with topological properties quite different from those of  BH can have similar observational properties from the perspective of a distant observer \cite{Virbhadra_Ellis, Gyulchev2019, Gyulchev2008AbdikamalovShaikh}.

Note that models of spherically symmetric compact objects with linear SFs are best studied for  massless linear SF, starting from the early publications \cite{Fisher, Janis,WymanVirbhadra}; more complicated  SF potentials have been considered elsewhere (see, e.g., \cite{ CruzLiebling, Zloshchastiev, Stashko}). As we will see below, main   asymptotic properties of the solutions are common for a fairly wide class of the SF potentials.

One of main issues of the present study  concerns the domain  of regular solutions and occurrence of singularities. Indeed, the singular behavior is very common for nonlinear differential equations and it is not evident that there is no "spherical" singularities outside the center (that is, for $r>0$ in the Schwarzschild coordinates). In particular,  equations of a static self-interacting  SF in special relativity (i.e. on the {\it fixed} Minkowsky background) can have  solutions with singularities at arbitrary spatial points (see Appendix \ref{SF_in_flat_space}).

{In case of linear scalar fields, the singularities outside the center of spherically symmetric configurations are not expected (cf.  \cite{Fisher, AsanovStashko3}). The question arises in case of non-linear SF models in view of the above special relativistic examples. Nevertheless, a general relativistic analysis  in the presence of nonlinear SFs  shows \cite{Stashko4} that such "spherical" singularities  do not seem to appear, if gravitational effects are involved; namely, solutions of the Einstein-SF equations are singular for $r\to 0$, but they do not have singularities for $r>0$. However, our paper\cite{Stashko4} uses numerical integration in case of particular SF potentials, so the  these results are not too general.  
	
In the present paper we give  a rigorous proof showing that, indeed, the joint system of Einstein-SF equations does not lead to "spherical" singularities for $r>0$, and we derive asymptotic relations for $r\to 0$  under  general conditions for positive definite SF potentials. Moreover, our results  extend to the case of several SFs with similar qualitative  metric behavior. Therefore, we state our findings for 
 asymptotically flat  space-times  in presence of $N$ nonlinear SFs, which represent static spherically symmetric configurations of General Relativity. The scalar fields are assumed to be minimally coupled to gravity.} 

In Section \ref{general} we formulate general requirements for the SF potentials {that are used to guarantee some properties of the solutions stated below.} The potentials are assumed to be exponentially bounded and to fulfill conditions analogous to that used in the proof of the Bekenstein \cite{Bekenstein} theorems. In Section \ref{section_qualitative} we present a direct proof  that  solutions of the joint system of Einstein -- SF equations obeying  conditions at spatial infinity are regular for all values of $r>0$ in the Schwarzschild (curvature) coordinates  except naked singularities in the center $r=0$.    In Section \ref{Asymptotics_center} we prove asymptotic relations for the metric and SF near the center. They are used in Section \ref{Section_Test_particle_motion} to describe  qualitative features of  the space-time geodesics, which are important for understanding the configuration image from the perspective of a remote observer. The main findings are summarized in Section \ref{Section_discussion}. In Appendix \ref{SF_in_flat_space} we give an example of a non-linear SF in the  flat space (static spherically symmetric case), showing the  appearance  of  singularities for $r>0$. Appendix \ref{N_free scalar fields} describes the generalized Fisher solution for $N$ free scalar fields. The results are illustrated in detail by an example of numerical solutions with one monomial SF  (Appendix \ref{Section_asymptotic+numerical}).

\section{Basic relations}
\label{general}
We consider $N$  real  scalar fields  $\Phi=\{\phi_1,...,\phi_N\}$ that are described by  Lagrangian density
\begin{equation}\label{lagrangian}
L=\frac{1}{2}\sum\limits_{i=1}^{N}\partial_{\mu}\phi_i\partial^{\mu}\phi_i-V(\Phi).
\end{equation}
Throughout the paper we assume that $V(\Phi)$ is a twice  continuously-differentiable function,
 \begin{equation}
 \label{positive potential_0}
 V(\Phi)\ge 0,
 \end{equation}
 and
 \begin{equation}
 \label{positive potential_dash}
 \phi_i V'_i(\Phi) \ge 0 ,\quad i=1,...,N,
 \end{equation}
 where   $V'_i= \partial V/\partial \phi_i$.

Assumptions (\ref{positive potential_0}, \ref{positive potential_dash}) can be fulfilled, e.g., in case of a polynomial potential
\begin{equation}
\label{example_V}
V(\Phi)=\sum_{n_{i_1}+...+n_{i_N}\ge 1}^{} w_{i_1,...i_N} \prod_{i=1}^{N} \phi^{2n_i}_i
\end{equation}
where  $w_{i_1,...i_N}\ge 0$.

Also we assume that there exist positive constants $C_0,\, C'_0, \,\kappa,\, \kappa'$ such that for all $\Phi$:
 \begin{equation}
 	\label{estimate_of_V0}
 	|V(\Phi)|<C_0 \exp(\kappa\norm{\Phi}),
 \end{equation}
{ ($\norm{\Phi}$ stands for the Euclidean  norm of the $N$-component vector $\Phi$)} and for all $i=1,2,...,N$
 \begin{equation}
 	\label{estimate_of_V}
 	\left|\frac{\partial V}{\partial \phi_i}\right| <C'_0 \exp(\kappa'\norm{\Phi}) .
 \end{equation}
Evidently, estimates (\ref{estimate_of_V0}, \ref{estimate_of_V}) are fulfilled for any finite degree polynomial including  example (\ref{example_V}). One can show,  if  (\ref{estimate_of_V}) is valid,  then  there exist some constants $C_0, \kappa $ such that  (\ref{estimate_of_V0}) is also valid; i.e. in fact only  (\ref{estimate_of_V}) is necessary.

The space-time endowed with the metric $g_{\mu\nu}$ subject to the Einstein equations\footnote{Units: G=c=1}
\begin{equation}\label{Ein_all}
G_{\mu}^{\nu}=8\pi T_{\mu}^{\nu}\,
\end{equation}
is assumed to be asymptotically flat.
The field equations
\begin{equation}
\label{dalembert}
g^{\mu\nu}\nabla_{\mu}\nabla_{\nu} \phi_i=-V'_i (\Phi), \quad i=1,...,N;
\end{equation}
follow  from  (\ref{lagrangian}).

The  energy-momentum tensor of the scalar fields is
\begin{equation}
\label{sf-e-m-tensor}
T_{\mu \nu } =  \sum\limits_{i=1}^{N}\partial_{\mu} \phi_{i }\partial_{\nu}\phi_{i} - g_{\mu \nu } L \, .
\end{equation}
We work with a static spherically symmetric space-time metric in curvature coordinates
\begin{equation}
\label{metric}
ds^2 = e^{\alpha(r)}dt^2 - e^{\beta(r)}dr^2 - r^2 dO^2,
\end{equation}
where  $dO^2=d\theta^2+(\sin\theta)^2 d\varphi^2$;  radial variable $r> 0$.

In case of   metric (\ref{metric}), the Einstein   equations
 yield
 \begin{equation}\label{Ein_1-0}
 	\frac{d}{dr} \left[ r \left(e^{-\beta}-1\right) \right]=-8\pi r^2 T_0^0 \,,
 \end{equation}
where $ T_0^0= e^{-\beta}\sum\limits_{i=1}^{N}\phi_i'^2/2+ V(\Phi)$ , $\phi'_i=d\phi_i/dr$,
  \begin{equation}
 	\label{Ein_2-0}
 	r e^{-\beta}\frac{d\alpha}{dr}+e^{-\beta}-1 =-8\pi r^2 T_1^1,
 \end{equation}
where $ T_1^1= -e^{-\beta}\sum\limits_{i=1}^{N}\phi_i'^2/2+ V(\Phi)$.

Equations (\ref{dalembert}) yield
 \begin{equation}
 \label{equation-phi}
\frac{d}{dr}\left[r^2 e^{\frac{\alpha-\beta}{2}}\frac{d\phi_i}{dr}\right]=r^2 e^{\frac{\alpha+\beta}{2}}V'_i(\Phi)\, ,
 \end{equation}
$i=1,...,N$.

 In view of the asymptotic flatness we assume
 \begin{equation}  \label{limits_0}
 \lim\limits_{r\to \infty}\left[ r(e^{\alpha}-1)\right]= \lim\limits_{r\to \infty}\left[ r(e^{-\beta}-1)\right] =-r_g,
 \end{equation}
 where $r_g=2M$ and $M>0$ is the configuration mass.

 It can be assumed that  at spatial infinity the SF components  behave as independent fields in the flat space and tend to zero.  We assume  $\Phi(r)\to 0$ for $r\to\infty$ and
  \begin{equation} \label{infinity_limit_phi1}
\exists K:\quad  r^2 \norm{\Phi'(r)}<K<\infty,
  \end{equation}
  whence also
    \begin{equation} \label{infinity_limit_phi}
    r \norm{\Phi(r)}<K \, .
    \end{equation}
  Stronger restrictions for SF can be assumed for some  potentials (see, Appendix \ref{SF_in_flat_space}), but conditions   (\ref{infinity_limit_phi1}, \ref{infinity_limit_phi}) are sufficient for our purposes.

  {\it Definition}. Functions  $\alpha(r), \beta(r) \in C^1$ and $\Phi(r)\in C^2$ will be said to be a solution of equations (\ref{Ein_1-0} -- \ref{equation-phi}) on $(r_0, \infty)$, $r_0\ge 0$, if they satisfy these  equations  on  $(r_0, \infty)$ and conditions (\ref{limits_0}, \ref{infinity_limit_phi1}, \ref{infinity_limit_phi}).

  Equations (\ref{Ein_1-0}, \ref{Ein_2-0})  are equivalent to
  \begin{equation}
  \label{Ein_A}
  \alpha' +\beta'=8\pi r \sum\limits_{i=1}^{N} \phi_i'^2\,,
  \end{equation}
  \begin{equation}
  \label{Ein_B}
  \beta'-\alpha'= \frac{2}{r}+e^{\beta}\left[16\pi r V(\Phi)-\frac{2}{r}\right] .
  \end{equation}
  Following \cite{Fisher}, instead of $\alpha$ and $\beta$ we introduce new (positive)  variables
  \begin{equation}\label{X_Y}
  X=e^{(\alpha+\beta)/2},\quad Y =r e^{(\alpha-\beta)/2}\, ,
  \end{equation}
  satisfying, in view of (\ref{limits_0}),
  \begin{equation}  \label{limits_XY}
  \lim\limits_{r\to \infty}\left[ r(X-1)\right]=0,\quad \lim\limits_{r\to \infty}\left( Y-r\right) =-r_g\,.
  \end{equation}

  Also we introduce
  \begin{equation}\label{Z_i}
  Z_i=-r Y \phi_i',\quad i=1,...,N .
  \end{equation}
  Conditions (\ref{infinity_limit_phi},  \ref{infinity_limit_phi1}) and the second condition of (\ref{limits_XY}) yield
   \begin{equation}  \label{limit_Z_i}
 |Z_i(r)|<K,\quad  \lim\limits_{r\to \infty} \left[\phi_i(r)Z_i(r)\right]=0.
   \end{equation}
  After simple transformations from (\ref{Ein_A}, \ref{Ein_B}) we get equivalent system
  \begin{equation}
  \label{Ein_A_2}
  \frac{dX}{dr}=  4\pi \frac{X }{r Y^2}\sum\limits_{i=1}^{N} Z_i^2 \,,
  \end{equation}
   \begin{equation}
   \label{Ein_B_2}
   \frac{dY}{dr}=X\left[1-8\pi r^2 V(\Phi)\right]\,.
   \end{equation}
   Equation  (\ref{equation-phi}) is reduced to a pair of the first-order equations
    \begin{equation}
    \label{equation-phi_2}
    \frac{dZ_i}{dr}= - r^2 X V'_i\,,
    \end{equation}
 $i=1,...,N$,
    \begin{equation}
    \label{equation-phi_2a}
    \frac{d\phi_i}{dr}= - \frac{Z_i}{rY }\,.
    \end{equation}

 \section{Regularity of solutions for $r> 0$} \label{section_qualitative}
As we mentioned above, for a nonlinear SF  it is necessary to be careful  about the  global behavior of solutions in connection with possible singularities that may arise when we continue the solutions from infinity to smaller values of the radial variable (see the example of   Appendix \ref{SF_in_flat_space} in case of the Minkowski space-time).

 In this Section we analyze the joint system  (\ref{Ein_1-0}, \ref{Ein_2-0}, \ref{equation-phi}) or equivalent system (\ref{Ein_A_2} -- \ref{equation-phi_2a}) and state some general conditions guaranteeing that the scalar field and the metric (\ref{metric}) is regular for all $r>0$. In order to do this one must rule out the cases $X(r)\to 0$, $Y(r)\to 0$, $|Z_i(r)|\to \infty$, $ |\phi_i(r)|\to\infty$ for $r\to r_0+0$ for some $r_0>0$.

 Below we use monotonicity properties of solutions following from  condition  (\ref{positive potential_dash}).
 Using (\ref{X_Y}, \ref{Z_i}, \ref{equation-phi_2}, \ref{equation-phi_2a}),  we get
\begin{equation}\label{phi-phi'}
-\frac{d}{dr}\left(\phi_i Z_i \right)=\frac{Z_i^2}{rY} + r^2 X\phi_i V'_i .
\end{equation}
{ Below we use  (\ref{X_Y}, \ref{Z_i}) to define $Z_i$.}

{\it Lemma}  1.   Let  condition  (\ref{positive potential_dash}) is valid for all $\Phi$, $\alpha(r),\, \beta(r)$ are continuously differentiable on $(r_0, \infty)$, $r_0\ge 0,$ and satisfy (\ref{limits_0}). Let, {for some $i$,} $\phi_i(r)\in C^2 $ is a non-trivial  solution of  (\ref{equation-phi}) on this interval {\bf ($\phi_i(r)\not\equiv 0$)} with conditions (\ref{infinity_limit_phi1}, \ref{infinity_limit_phi}). Then functions
$\phi_i(r)$,  $Z_i(r)$ and  $d\phi_i/dr$ do not  change their signs, $\phi_i(r)Z_i(r)>0$ and 
$\phi_i(r)d\phi_i/dr< 0$ on  $(r_0,\infty)$.

{\it Proof.}  {
Let $r_1$ be an arbitrary point to the right of $r_0$. The r.h.s.  of  (\ref{phi-phi'}) is non-negative in view of (\ref{positive potential_dash}) and $\phi_i Z_i$ is non-increasing.  If we suppose (on the contrary) that $\phi_i(r_1)Z_i(r_1)<0$ for  $r_1>r_0$, then this will be preserved for $r>r_1$ in contradiction to  $\phi_i(\infty) Z_i(\infty)=0$, which is the consequence of  (\ref{limits_0}, \ref{infinity_limit_phi1},  \ref{infinity_limit_phi}). Therefore,  $\phi_i(r)Z_i(r)\ge 0$ for  $r>r_0$. Now, if we suppose $\phi_i(r_1)Z_i(r_1)=0$ (again on the contrary), then  we have analogously that $\phi_i(r)Z_i(r)\equiv 0$ that is $\phi_i(r)d\phi_i/dr\equiv 0$ and $d\phi^2_i/dr\equiv 0$ for $r>r_1$. In view of $\phi_i(\infty)=0$ this yields $\phi_i(r)\equiv 0$ for $r>r_1$ in contradiction to the assumption that $\phi_i(r)$ is non-trivial.
This yields strict inequality $\phi_i(r)Z_i(r)>0$; then $\phi_i(r)$ and  $Z_i(r)$ cannot change their signs.} This proves all the statements of this {\it Lemma}.
\vspace{\baselineskip}

Further we assume that at least for one component of $\Phi$ is non-trivial: $\phi_{i}(r)\ne 0$.

{\it Lemma} 2. Let  conditions  (\ref{positive potential_0}, \ref{positive potential_dash}) are fulfilled,   functions $\alpha(r),\,\beta(r),\,\Phi(r)\in C^1$  satisfy equations (\ref{equation-phi}, \ref{Ein_A}, \ref{Ein_B})  and  $\phi_{i}(r)\ne 0$ for $i=1,...,N$ in $(r_0,r_1]$, where $0<r_0<r_1<\infty$.
Then there exists  $\eta_0>0:$ $Y(r)>\eta_0$ and $S_i Z_i(r)>S_i Z_i(r_1)>0$, where $S_i= \mathrm{sign}\,\phi_i$.

{\it Proof.} We use system (\ref{Ein_A_2} -- \ref{equation-phi_2a}).

Let for some $i$ we have $\phi_{i}(r)\ne 0$. In view of {\it Lemma} 1 we can assume $\phi_{i}(r)>0$, $Z_{i}(r)>0$, $\phi'_{i}(r)<0$  without loss of generality. Then  $\phi_{i}(r)$ is monotonically decreasing. In view of (\ref{equation-phi_2}) and (\ref{positive potential_dash}),
 $Z_{i}(r)>0$ is decreasing and $Z_{i}(r)>Z_{i}(r_1)$ for $r<r_1$. Analogously, inequality $S_i Z_i(r)>S_i Z_i(r_1)$ is fulfilled for the other non-trivial SF components.

 In view of (\ref{Ein_A_2}), function $X(r)$ is monotonically increasing.
From (\ref{Ein_A_2}, \ref{Ein_B_2}) we have for $r<r_1$
\begin{equation*}
\label{dY/dX}
\frac{1}{Y^2} \frac{dY}{dX}=\frac{r}{4\pi\sum\limits_{i=1}^{N} Z_i^2}\left[1-8\pi r^2 V(\Phi)\right]\le
\end{equation*}
(using (\ref{positive potential_0}))
\[\le \frac{r_1 }{4\pi\sum\limits_{i=1}^{N} Z_i^2(r)}\le
\frac{r_1 }{4\pi\sum\limits_{i=1}^{N} Z_i^2(r_1)},
\]
where   we take into account $X'>0$, and we used the monotonicity properties of $S_i Z_i$.
Integration of this inequality yields
\begin{equation*}
\frac{1}{Y(r)}\le\frac{1}{Y_1}+  \frac{r_1 X_1}{4\pi\sum\limits_{i=1}^{N} Z_i^2(r_1)}
\end{equation*}
Therefore, $1/Y(r)>0$ is bounded and $\exists \eta_0>0:$ $Y{(r)}>\eta_0$.
The {\it Lemma} 2 is proved.
\vspace{\baselineskip}

On account of this  {\it Lemma} we see that if there is a non-trivial component  $\phi_i(r)\ne 0$  that satisfies equations (\ref{equation-phi}, \ref{Ein_A}, \ref{Ein_B})  in $(r_0,r_1]$, where $0<r_0<r_1<\infty$, then there exists  $\eta_0>0:$ $Y(r)>\eta_0$.

 {\it Lemma} 3.   Let the conditions (\ref{estimate_of_V0}, \ref{estimate_of_V}) are fulfilled and   functions $\alpha(r),\,\beta(r) \in C^1$,  $\Phi(r)\in C^2$, $\phi_i\ne 0$ (at least for some $i$) satisfy equations (\ref{Ein_A}, \ref{Ein_B}) and (\ref{equation-phi})  on    $(r_0,r_1]$, where $0<r_0<r_1<\infty$. Then  there exist finite limits
 \begin{equation} \label{limits_lemma_2}
 \bar Y(r_0)=\lim\limits_{r\to r_0+0}Y(r)> 0, \,\,\bar Z_i(r_0)=\lim\limits_{r\to r_0+0}Z_i(r)>0,
 \end{equation}
  \begin{equation} \label{limits_lemma_3}
  \bar X(r_0)=\lim\limits_{r\to r_0+0}X(r)> 0, \,\,\bar \phi_i(r_0)=\lim\limits_{r\to r_0+0}\phi_i(r)\ne 0.
  \end{equation}

   {\it Proof.}
    According to the assumption of this {\it Lemma},  $X(r_1)$ is finite.  Let  $r_0<r\le r_1$. Equation (\ref{Ein_A}) on account of (\ref{X_Y}) yields
   \begin{equation} \label{eq2 lemma}
   X(r)=X(r_1)\exp{-4\pi \int\limits_{r}^{r_1}x\sum\limits_{i=1}^{N}\phi_i'^2(x)dx}
   \end{equation}
   Using the Cauchy -- Bunyakovsky -- Schwarz inequality we have for $r<r_1$
   \[
   |\phi_i(r)-\phi_i(r_1)|=\left|\int\limits_{r}^{r_1}(\phi_i'(x)\sqrt{x}\,)\cdot \frac{1}{\sqrt{x}}\cdot dx \right|\le
   \]\[
   \le \int\limits_{r}^{r_1}\left|\phi_i'(x)\right|\sqrt{x}\,\cdot \frac{1}{\sqrt{x}}\cdot dx \le
   \sqrt{\int\limits_{r}^{r_1}x[\phi_i'(x)]^2dx\, \ln(r_1/r)}\,\,.
   \]
   Then
   \begin{equation} \label{lemma2_A}
   \int\limits_{r}^{r_1}x[\phi_i'(x)]^2dx\ge\frac{[\phi_i(r)-\phi_i(r_1)]^2}{\ln(r_1/r)},
   \end{equation}
   whence using (\ref{eq2 lemma}) we have
    \begin{equation}
    X(r)\le X(r_1) \exp\left\{ -{4\pi}\sum\limits_{i=1}^{N}\frac{\left[ {\phi_i(r)-\phi_i(r_1)}\right]^2}{\ln(r_1/r)}\right\}, \label{estimate_Lemma_1}
    \end{equation}
 and we strengthen this inequality    by replacing $\ln(r_1/r)$ by $\ln(r_1/r_0)$:
\begin{equation}
X(r)\le X(r_1) \exp\left\{ -{4\pi}\sum\limits_{i=1}^{N}\frac{\left[ {\phi_i(r)-\phi_i(r_1)}\right]^2}{\ln(r_1/r_0)}\right\}. \label{estimate_Theorem_1}
\end{equation}
Denote
\begin{equation} \label{b(r)}
B(r)=	X(r)|V(\Phi(r))|\,,\quad  \tilde B_i(r)=	X(r)|V_i'(\Phi(r))| .
\end{equation}
  As $\norm{\Phi}\le \sum\limits_{i=1}^{N}|\phi_i|$, then according to (\ref{estimate_Theorem_1}) and (\ref{estimate_of_V0}) we obtain
 \flushleft $B(r) \le $
 \begin{equation}
 	\le C_1 \exp\left\{ \sum\limits_{i=1}^{N}\left[ -4\pi\frac{[\phi_i(r)-\phi_i(r_1)]^2}{\ln(r_1/r_0)}+
 	\kappa|\phi_i(r)| \right]\right\}\, \label{lemma2_B}
 \end{equation}
 where $C_1=X(r_1)C_0>0$.
 Term $\sim 4\pi |\phi(r)|^2/\ln(r_1/r_0)$  dominates the exponent for $\phi\to \infty$, the expression in the exponent as a function of $\phi$  has maximum, so $B(r)$ is uniformly bounded for $r\to r_0+0$, $r_0>0$.
{   Analogous consideration shows that $\tilde B_i(r)$ is also bounded} (even if $\phi_i\to \infty$). Then expressions (\ref{b(r)}) and  the right-hand sides of (\ref{Ein_B_2}, \ref{equation-phi_2}) are bounded, integrable yielding the existence of limits $\bar Y(r_0),\bar Z_i(r_0)$. Inequalities $\bar Y(r_0)>0$, $S_i\bar Z_i(r_0)>0$ follow from considerations of {\it Lemmas} 1,2.

 From (\ref{equation-phi_2a}) in view of {\it Lemma} 1 follows that $|d\phi_i/dr|$ and $\phi_i(r)$ are bounded and have limits for $r\to r_0$. Existence of $\bar X(r_0)>0$ follows either from (\ref{Ein_A_2}) or directly from (\ref{Ein_A}) in view of the previous  results. The {\it Lemma} 3 is proved.
 \vspace{\baselineskip}

\setlength\parindent{14pt}
We summarize the above statements in the form of the following

 {\it  Theorem}. Let the SF potential satisfies conditions (\ref{positive potential_0},  \ref{positive potential_dash}) and (\ref{estimate_of_V0}, \ref{estimate_of_V})   for all $\Phi$. Let   $\alpha(r),\,\beta(r),\in C^1$, $\Phi(r)\in C^2$   represent a non-trivial ($\phi_i(r)\not\equiv 0$, $i=1,...,N$) solution of  equations (\ref{equation-phi}, \ref{Ein_A}, \ref{Ein_B}) on open interval $(r_0,\infty), \,r_0>0$ with conditions (\ref{limits_0}, \ref{infinity_limit_phi1},  \ref{infinity_limit_phi}). Then

 (i) there exist finite limits of functions $\alpha(r),\,\beta(r),\, \phi_i(r)$ and $\phi'_i(r)$ for $r\to r_0$;

 (ii) this solution can be regularly continued  onto a left  neighborhood of $r_0$;

(iii)  this solution can be regularly continued   for all $r>0$ up to the center.

{\it Proof.} Statement (i) of the theorem is essentially the result of {\it Lemma} 3. The right hand sides of equations (\ref{Ein_A_2} -- \ref{equation-phi_2a}) are analytic in the neighborhood of $\bar X(r_0)>0$, $\bar Y(r_0)>0$, $S_i\bar Z_i(r_0)>0$, $S_i\bar \phi_i(r_0)>0$. Then statement (ii) follows from the existence-uniqueness theorem for ordinary differential equations.
Application of the continuous induction in order to continue the solutions for all $r>0$ completes the proof.

We note that the regularity for $r>0$ does not exclude a singularity at the origin $r=0$.

  \section{Asymptotics at the center} \label{Asymptotics_center}
  The next question concerns the behavior of the solutions in the vicinity of the center that can be studied using considerations similar to {\it Lemma} 3 with some restrictions on $\kappa, \kappa'$ from (\ref{estimate_of_V0}, \ref{estimate_of_V}). We take into account that signs $S_i\equiv \mathrm{sign}(\phi_i)= \mathrm{sign}  (Z_i)$ do not change on $(0,\infty)$.

{\it Lemma} 4. Let conditions (\ref{positive potential_0},  \ref{positive potential_dash}) and (\ref{estimate_of_V0}, \ref{estimate_of_V})  are fulfilled with $\max {(\kappa^2, \kappa'^2)}<32\pi/N$. Let $\alpha(r), \beta(r), \phi_i(r)\not\equiv 0$ $(i=1,...,N)$ represent a solution of  (\ref{Ein_1-0}, \ref{Ein_2-0}, \ref{equation-phi}) on $(0, \infty)$ with conditions (\ref{limits_0}, \ref{infinity_limit_phi1},  \ref{infinity_limit_phi}). Then there exist finite nonzero limits
  \begin{equation}\label{Q more 0}
Z_{i,0}=\lim\limits_{r\to 0+0} Z_i(r), \quad Y_0=\lim\limits_{r\to 0+0} Y(r)
  \end{equation}
such that $S_iZ_{i,0}>0, \,Y_0>0$.

{\it Proof.}    Let  $0<r<r_1<\infty$. { We denote $L=\ln(r_1/r)$ and $D(r)= r^2B(r)$ that appears in r.h.s. of (\ref{Ein_B_2})}.
Now we repeat considerations of {\it Lemma} 3 leading to  (\ref{lemma2_B}), but we can leave $L=\ln(r_1/r)$ in this inequality instead of $\ln(r_1/r_0)$ {yielding:
 \flushleft $D(r)=r_1^2 (r/r_1)^2B(r)=r_1^2 e^{-2L}B(r)\le C_2   e^{-2L}\cdot$}
 \begin{equation}
 \cdot\exp\left\{-\sum\limits_{i=1}^{N}\left[ \frac{4\pi}{L}[\phi_i(r)-\phi_{i,1}]^2-
 \kappa|\phi_i(r)| \right]\right\}\, \label{lemma4_B}
 \end{equation}
 where $C_2=r_1^2 X(r_1)C_0>0 $, $\phi_{i,1}=\phi_i(r_1)$.

\setlength\parindent{14pt}
We strengthen the inequality by
discarding some negative terms in the exponent. After simple calculations {(completing a perfect square)} we get
 \[D(r)\le C_2\exp\left\{-2L+\frac{4\pi}{L}\sum\limits_{i=1}^{N}
  \left[\left(|\phi_{i,1}|+\frac{\kappa L}{8\pi}\right)^2\right]\right\} \]
If $\kappa^2<32\pi/N$, then this  expression is bounded for $r\to 0$ ($L\to \infty$); then the right-hand side of (\ref{Ein_B_2}) is integrable and the limit $Y_0\ge 0$ exists.

Analogously, under suppositions of this {\it Lemma} we get that $r^2 XV'_i$ in the right hand side of (\ref{equation-phi_2}) is integrable and limits $  Z_{i,0}$ exist.
  After that, inequalities $S_i Z_{i,0}>0, \,Y_0>0$ are obtained similarly to {\it Lemma} 3, Q.E.D.
  \vspace{\baselineskip}

Using {\it Lemma} 4 we can obtain the asymptotic behavior of SF for $r\to 0$. From (\ref{equation-phi_2a}) and using (\ref{Q more 0}) we have
\begin{equation}
\label{asymptotics_phi} \frac{d\phi_i}{dr}\sim  - \frac{\zeta_{i,0}}{r}, \quad
\phi_i(r)\sim -{\zeta_{i,0}}{\ln r}\,,
\end{equation}
where $\zeta_{i,0}=Z_{i,0}/Y_0$. The singularity of $\phi_i(r)$ for $r\to 0$ is a physical one; it takes place in any coordinate system and   cannot be removed by a coordinate transformation\footnote{{ Scalar curvature $R\sim-D_1/r^{\eta+3}$  and  Kretschmann scalar  $R_{\alpha \beta \gamma \delta}R^{\alpha \beta \gamma \delta}\sim  D_2/r^{2\eta+6}$ diverge for $r \to 0$, where constants $D_1>0, D_2>0$ depend on details of asymptotic behavior of $V(\Phi)$ for large fields.}}.

From (\ref{Ein_A}, \ref{Ein_B}) we obtain the leading terms of asymptotics for $r\to 0$:
\begin{equation}
\label{asymptotics_alpha_beta}
\alpha(r)\sim{(\eta-1)}{\ln r},\quad
\beta\sim{(\eta+1)}{\ln r},
\end{equation}
where $\eta=4\pi \sum\limits_{i=1}^{N}\zeta_{i,0}^2$.

Note that the asymptotics (\ref{asymptotics_phi}, \ref{asymptotics_alpha_beta}) are  similar to those  of the generalized Fisher solution with  $V\equiv 0$ (see Appendix \ref{N_free scalar fields}). Indeed,  under conditions (\ref{estimate_of_V0}, \ref{estimate_of_V}) the terms containing  $V(\phi)$ are asymptotically much smaller as compared with  the other terms in (\ref{Ein_A}, \ref{Ein_B}) for $r\to 0$.

\section{Test particle motion} \label{Section_Test_particle_motion}
The asymptotic relations (\ref{asymptotics_alpha_beta}) enable us to highlight main qualitative situations
concerning  the geodesic structure of the space-time around   the spherically symmetric static configuration with $N$ SFs.

 The integrals for trajectories of photons and test particles in the plane $\theta=\pi/2$ are
 \begin{equation}
 \label{geodesics_1}
 e^{\alpha}\left(\frac{dt}{d\tau}\right)^2 - e^{\beta}\left(\frac{dr}{d\tau}\right)^2  - r^2\left(\frac{d\varphi}{d\tau}\right)^2=S\,,
 \end{equation}
  \begin{equation}
 \label{geodesics_2}
 e^{\alpha}\left(\frac{dt}{d\tau}\right)=E, \quad  r^2\left(\frac{d\varphi}{d\tau}\right)=L\,,
 \end{equation}
 where $S=0$ in case of photons and $S=1$ for test particles, $\tau$ is a canonical parameter, $E>0$ and $L$ are constants of motion.
 This yields
 \begin{equation}
 \label{1D particle}
 e^{\alpha+\beta}\left(\frac{dr}{d\tau}\right)^2=E^2-
 U_{eff}(r,L,S)  \,,
 \end{equation}
 where $ U_{eff}(r,L,S)=e^{\alpha}\left(S+{L^2}/{r^2}\right)$.

Thus, we are dealing with  one-dimensional particle motion in the field of  effective potential $U_ {eff}$.

In view of asymptotics  (\ref{limits_0}), we have  $U_ {eff}\approx S+{L^2}/{r^2}$ for $r\to \infty$.
 In case of the radial motion of photons ($L=0$, $S=0$) using (\ref{asymptotics_alpha_beta}), for $r\to 0$ we have $r^{\eta}dr/d\tau\approx \pm E$, so photons can reach the singularity at the center for a finite value of $\tau$. For $L\ne 0$ both for $S=0$ and $S=1$,  we have two main situations defined by the sign of $\eta-3$. In Appendix \ref{Section_asymptotic+numerical}  we consider an example showing that both signs may be indeed possible.
  For $r\to 0$ and $L\ne 0$ asymptotic relations  (\ref{asymptotics_alpha_beta}) yield $U_ {eff}\approx L^2 e^{\alpha}{r^{-2}}\sim r^{\eta-3}$. Thus for $\eta>3$ we have $U_{eff}(r,L,S)\to 0$ for $r\to 0$ and for sufficiently large $L$ there is a maximum of $U_{eff}(r,L,S)$ as a function of $r$. Otherwise, for $\eta <3$ we have $U_{eff}(r,L,S)\to \infty$ for $r\to 0$.

Let us consider in more detail the motion of photons ($S=0$). In this case by an appropriate choice of the canonical parameter we can put $E=1$. { To build the image of a radiating accretion disk, the inverse ray tracing is widely used: instead of tracking the photons emitted by the disk, we track the trajectories of incident  photons moving in the opposite direction from the observer. For this purpose, we consider below scattering of photons  by the singularity in the center.}

Let $\eta>3$. Then there exists a global maximum of the effective potential
$\max {U_{eff}(r,L,0)}=L^2 M_0 $, where $M_0=\max{e^\alpha/r^2}$. The solutions $r(\tau)$ of (\ref{1D particle}) that describe incoming photons with $L^2M_0<1$, $dr/d\tau<0$, can be continued to the values $r\to 0$, that is, these photons  fall on the center. There is a non-zero capture cross section of the incident photons by the singularity  (see Fig. \ref{Ray-tracing_fall} of Appendix \ref{Section_asymptotic+numerical} for the example). On the other hand, if the singularity does not radiate, the external observer will see a dark spot in the center surrounded by a luminous ring due to radiating substance around the configuration.  The occurrence of the maximum of $\max {U_{eff}(r,L,0)}$  means that there exists at least one "photon sphere" (cf., e.g. \cite{Virbhadra_Ellis}) -- the set of unstable circular photon trajectories. The  situation with this configuration {  may be very similar to that for BH, when we  have  the same qualitative behavior of the test body circular orbits and the photon orbits (as described in Appendix \ref{Section_asymptotic+numerical}). In this case it will be difficult} to distinguish images of these objects without  additional independent information about the surrounding matter.

In case of $\eta<3$, $L\ne 0$ the effective potential $U_{eff}(r,L,S)$ is unbounded for $r\to 0$. Therefore, photons falling from infinity with $L\ne 0$ are reflected back from the  potential and do not reach the center. {  If the deflection angle $\alpha_d(L)$ is sufficiently large,  the incident photon  may not hit the disk plane at all. If, e.g., the disk is observed face-on, this will be the case of $\alpha_d(L)> \pi/2$ -- for sufficiently small impact parameters $L\ne 0$. The example of such a behavior is given in Fig. \ref{Ray-tracing_no_fall} of Appendix \ref{Section_asymptotic+numerical}. This means that the distant observer cannot receive photons with $L<L_0$ (from any part of the disk), that is he will see a dark spot in the center. This situation } is described in Appendix \ref{Section_asymptotic+numerical}, where we present a detailed consideration of null geodesics by the example of one non-linear SF with the monomial potential.

\section{Discussion}\label{Section_discussion}
We considered static spherically symmetric configurations of General Relativity in presence of $N$ scalar fields  minimally coupled to gravity, which obey conditions of asymptotic flatness. The SF potential is supposed  to satisfy conditions   guaranteeing a monotonic dependence of non-trivial SF modes upon radial variable  (\ref{positive potential_0}, \ref{positive potential_dash}); also  it must be exponentially bounded (\ref{estimate_of_V0}, \ref{estimate_of_V}). These conditions are fulfilled for a number of widely used field-theoretic models, such as
  positively definite polynomial potentials (\ref{example_V}). A superposition of independent SFs with monomial potentials $\sim \phi^{2n}_i$ can serve as a simple example.

Under these conditions, we proved that any asymptotically flat solution of the Einstein--SF equations cannot have singularities on a sphere of non-zero radius, i.e. the solutions are regular for all non-zero values $r>0$ of radial variable $r$ in the Schwarzschild (curvature) coordinates outside the center. On the other hand, all non-trivial SF  components $\phi_i$   have   singularities for   $r\to 0$. The results are illustrated by the numerical example for one SF with monomial potential (Appendix \ref{Section_asymptotic+numerical}).

We note that asymptotic properties of the solutions with $N$ non-linear scalar fields are remarkably similar to that of the case of one linear massless SF \cite{Fisher, Janis,WymanVirbhadra} and its generalization for $N$ free SFs (Appendix \ref{N_free scalar fields}). All these cases have naked singularities at the center with the logarithmic asymptotic behavior of SF and power-law metric components.

It would be interesting to relax the restrictions on the potentials (\ref{positive potential_0}, \ref{positive potential_dash}, \ref{estimate_of_V0}, \ref{estimate_of_V}). Though we note  examples (see, e.g. \cite{Zloshchastiev, Stashko}) showing that violation of the conditions  (\ref{positive potential_0}, \ref{positive potential_dash}) can lead to the black hole configurations  with scalar hair. One can also suppose that   non-linear SF potentials with a more strong dependence upon SF than in (\ref{estimate_of_V0}, \ref{estimate_of_V}) (like  $V(\phi)\sim \sinh(\phi^{2n}),  n>2$) can lead to spherical singularities with a non-zero radius.

In view of the asymptotic  properties of the metric we show that there exist two  different  types of the geodesic structure, depending on the strength of the SF components at infinity.  They are related to different behaviors of isotropic geodesics.
In case of the first one the incident photons with sufficiently  small impact parameters $L$ are captured by the singularity. There exists a "photon sphere" \cite{Virbhadra_Ellis} and spiral trajectories of photons passing near it that fall to the center after several revolutions.
This case is similar to that of BH,  and here one can also have the ring-like image of the accretion disk with a dark spot in the center. Different brightness distributions over the ring are possible, which, however, are also a function   of complicated physical processes that are unattainable for  independent observations.
Therefore, there is a lot of configurations with naked singularity that can mimic the Schwarzschild BH.

In the other type, the photons and the test particles with non-zero angular momentum cannot reach the center. The examples show that in this case the incident photons can be strongly deflected by the center, which makes it  impossible for some regions near the singularity to be observed from some directions at infinity; in this case the dark spot in the center can be possible as well.

\acknowledgments
We are grateful to anonymous referee for helpful remarks and suggestions. We thank  Yu.V.~Shtanov for fruitful discussions. This work has been supported by scientific
program "Astronomy and space physics" of Taras Shevchenko National University of Kyiv (project 19BF023-01).

\appendix

\section{One SF in the flat space}\label{SF_in_flat_space}
{\bf A1. Asymptotic behavior for $r\to\infty$.}
Here we consider one real SF $\phi$ with the power-law potential
\begin{equation}\label{[power-law]}
V(\phi)=w|\phi|^p, \quad w>0,
\end{equation}
The SF equation in  Minkowski space-time  is
 \begin{equation}
 \label{equation-phi_flat}
 \frac{d}{dr}\left[r^2 \frac{d\phi}{dr}\right]=pw r^2 \phi |\phi|^{p-2}.
 \end{equation}
We assume conditions (\ref{infinity_limit_phi1}, \ref{infinity_limit_phi}) to be fulfilled for $\Phi(r)\equiv \phi(r)$.
Due to considerations analogous to {\it Lemma} 1  of Section \ref{section_qualitative} we infer that $\phi(r)$ and $\phi'(r)$ do not change their signs and without loss of generality  we further assume $\phi>0, \,\phi'(r)<0$.

In case of $p=2$, $w=\mu^2/2$ (massive SF with mass $\mu$) there is the exact solution satisfying zero condition for $r\to\infty$:
  \begin{equation}  \label{phi_flat_n2}
   \psi(r)=\frac{Q}{r}\exp(-\mu r),\quad Q=const.
  \end{equation}

For $p>2$, substitution
  \begin{equation}\label{substitution}
 \phi=e^{-qt}\psi,\quad t=\ln r,\quad q=\frac{2}{p-2}
   \end{equation}
 leads to   autonomous differential equation
  \begin{equation}  \label{equation-phi_flat_t2}
  \frac{d^2\psi}{dt^2}+(-2q+1) \frac{d\psi}{dt}+q(q-1)\psi=pw \psi^{p-1}.
  \end{equation}
 Equation (\ref{equation-phi_flat_t2}) is equivalent to the dynamical system on the plane
 \begin{equation}\label{2-dynamical}
  \frac{du}{dt}=(2q-1) -q(q-1)\psi+pw \psi^{p-1},\quad  \frac{d\psi}{dt}=u .
 \end{equation}\\

 (i) First consider the case  $2<p<4$  when $q(q-1)=2(4-p)(p-2)^{-2}>0$. On the half-plane $\psi\ge 0$  the dynamical system  (\ref{2-dynamical})  has critical points $(\psi=0,\,u=0)$  and  $(\psi= Q_0,\,u=0)$, where
\begin{equation}\label{Q_0}
Q_0=\left[\frac{q(q-1)}{pw}\right]^{\frac{1}{p-2}} \,.
\end{equation}
The point (0,0) is a repeller (unstable node) with the eigenvalues of the linearized  system
\begin{equation}\label{lambda3}
\lambda^*_1=q=\frac{2}{p-2},\quad \lambda^*_2=q-1=\frac{4-p}{p-2}.
\end{equation}
The point $(Q_0, 0)$ is a saddle with eigenvalues
\begin{equation}\label{lambda1}
{\lambda}_{\pm}=\frac{6-p}{2(p-2)}\left[1\pm \sqrt{ 1
	+\frac{8(4-p)(p-2)}{(6-p)^2}}\right],
\end{equation}
The separatix branches that enter the saddle correspond to asymptotic solutions of (\ref{equation-phi_flat}) for $r\to\infty$. For these branches we have
\[\psi(t)\approx Q_0[1+C\exp(-\lambda t)], \quad t\to \infty,\]
 where $C$ is an arbitrary constant, $\lambda=-\lambda_{-}>0$. Correspondingly, we have asymptotic solution of (\ref{equation-phi_flat}) for $r\to \infty$:
 \begin{equation}\label{phi_1(r)}
 \phi_1(r)\approx \phi_1(r)= \frac{Q_0}{r^q}\left(1+\frac{C}{r^{\lambda}}\right)\,  ,
 \end{equation}
 The other solutions near the saddle do not fit condition $\phi(\infty)= 0$.
\newline

(ii) For $p>4$  there is the only critical point (0,0) and this is a saddle   (see (\ref{lambda3}) for $p>4$).  Condition $\phi(\infty)= 0$ leads to the separatrix of the saddle yielding $\psi(t)\sim Q e^{-|\lambda_2^*| t}, \,t\to \infty$ and
 \begin{equation}
 \phi(r)\approx   Q /r,\, r\to \infty, \label{flat_phi_more4}
 \end{equation}
$Q$ being an arbitrary constant.
\newline

(iii) For $p=4$ $(q=1)$ there is also the only critical point (0,0), but it is not simple. Analogously to the case $p>4$, there are  solutions that tend to (0,0) for $t\to \infty$ yielding  asymptotic  solution for SF
 \begin{equation}\label{phi_inf_4}
 \phi(r)=\frac{Q}{r\sqrt{|\ln r|}}\left(1+\frac{3\ln|\ln r|}{4\ln r}+...\right)\,
 \end{equation}
 where $Q$ being an arbitrary constant.
 \newline

{\bf A2. Spherical singularities.}
Typical phenomenon for some types of a non-linear equation is the occurrence of singularities for finite values of the independent variable. Here we present an example showing that this may be the case for certain solutions of equation (\ref{equation-phi_flat}).

First of all, one can check directly that  with   $p>2$ there exists a solution of equation (\ref{equation-phi_flat}), which is singular at $r=r_s>0$, which can be  represented approximately near this point as
\[
\phi(r)\approx\left[\frac{q(q+1)}{pw(r-r_s)^{2}}\right]^{q/2}
\]
 with arbitrary $r_s>0$ and $q=2/(p-2)$. However, here we do not know the behavior of this solution for $r\to\infty$.
In this view we present an example with more detailed consideration of the solutions satisfying asymptotic  condition (\ref{flat_phi_more4}) at infinity. We confine ourselves to the case $p=2n,\,\,n>2$.

\begin{figure}[h]
	\vskip1mm
	\includegraphics[width=80mm]{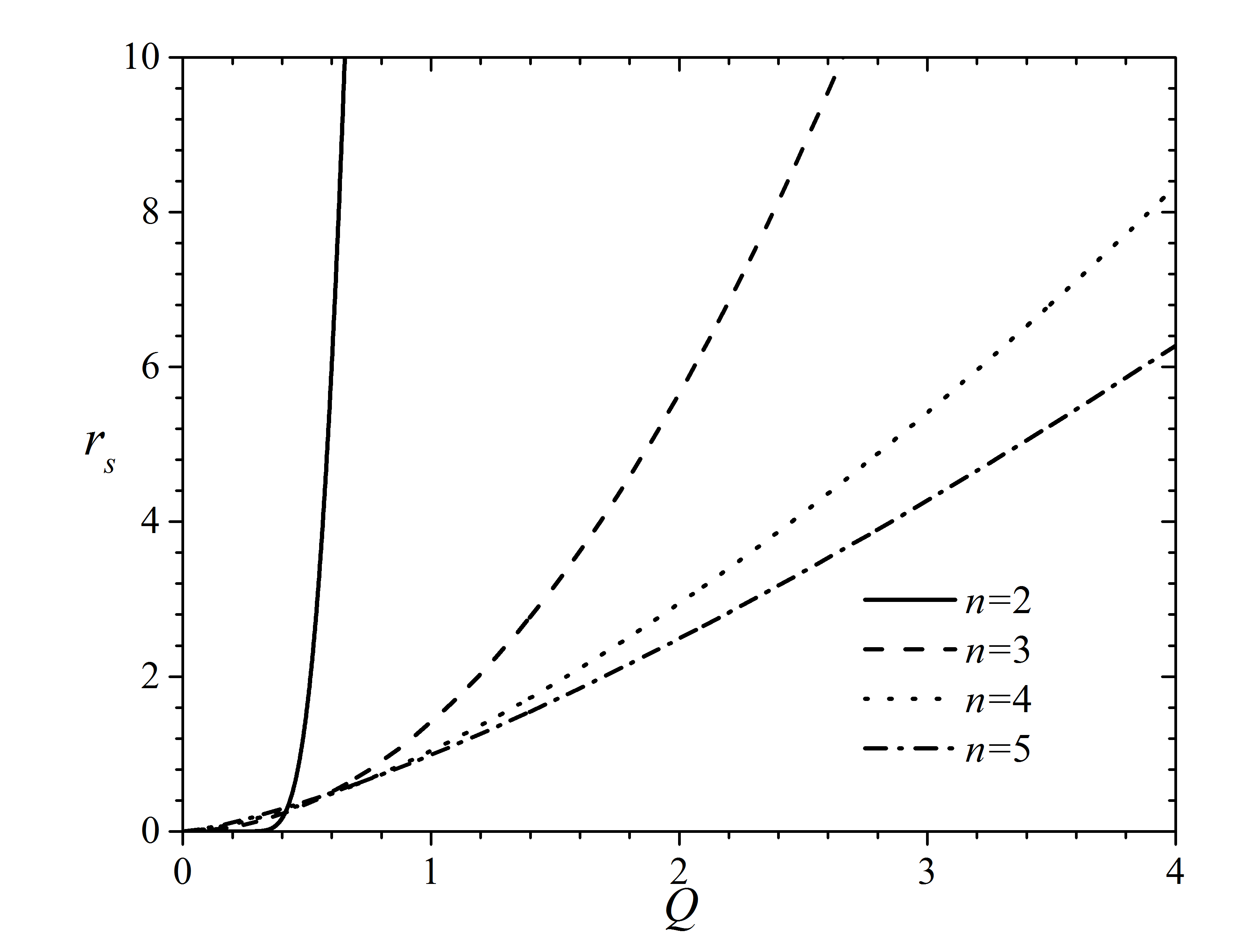}
	\vskip-3mm\caption{Position of singularity $r_s$ for the solution of (\ref{equation-phi_flat}) with different $p=2n$  as a function of $Q$ from conditions (\ref{flat_phi_more4}, \ref{phi_inf_4}).}\label{r_s_of_Q2}
\end{figure}

Multiplying (\ref{equation-phi_flat}) by $\phi'(r)$ after some transformations we get
 \begin{equation}\label{dEdr}
   \frac{dE}{dr} =-\frac{2}{r} [\phi'(r)]^2\le 0,
 \end{equation}
where $E(r)=[\phi'(r)]^2/2-w\phi^{2n}$. Then for $r<r_0$
we have $E(r)\ge E(r_0)$. For a sufficiently large $r_0$ we have
 \begin{equation}\label{asympt_phi}
\phi(r_0)\approx Q/r_0<<1,\quad  \phi'(r)\approx -Q/r_0^2,
 \end{equation}
 this yields $E(r_0)> 0$ and
 \begin{equation*}
[\phi'(r)]^2 >w\phi^{2n}(r).
 \end{equation*}
Taking into account the signs ($\phi>0, \,\phi'<0$)
 \begin{equation*}
 \phi'(r) <- \sqrt{w}\phi^{n} \quad\to\quad \frac{d}{dr}\left(\frac{1}{\phi^{n-1}}\right)> \sqrt{w}(n-1).
 \end{equation*}
Integration of this inequality  yields  on $[r,r_0]$, ($r<r_0$)
 \begin{equation*}
 \phi(r) > \left[ \phi_0^{-(n-1)}-\sqrt{w}(n-1)(r_0-r)\right]^{1/(n-1)} .
 \end{equation*}
If
\begin{equation} \label{singularity}
 \phi_0^{-(n-1)}-\sqrt{w}(n-1)r_0<0\,,
\end{equation}
then we necessarily have singularity of $\phi$ for some $r=r_s>0$.
The inequality (\ref{singularity})  needs to be checked to be compatible with
(\ref{asympt_phi}). Both estimates will be satisfied for a sufficiently large $r_0$  and
\begin{equation*}
Q>r_0^{(n-2)/(n-1)}[\sqrt{w}(n-1)]^{-1/(n-1)} .
\end{equation*}
This ensures the existence of the singularity for some $r=r_s>0$. However, this seems to be a too tight assessment.  Numerical integration shows that the singularity occurs for much lower $Q$ (see Fig. \ref{r_s_of_Q2}).

  \section{Generalized Fisher solution: $N$ free scalar fields} \label{N_free scalar fields}
Here we consider the case of $N$ scalar fields with $V(\Phi)\equiv 0$. Our considerations  closely follow the Fisher work \cite{Fisher}.

From (\ref{equation-phi_2}) we have $Z_i\equiv const$, and (\ref{Ein_A_2}, \ref{Ein_B_2}) are separated yielding the second order equation
\begin{equation}\label{Fisher_1}
\frac{d^2Y}{dr^2}= \frac{\Xi}{rY^2}\frac{dY}{dr}\,,
\end{equation}
where
\begin{equation}\label{Ksi}
\Xi=4\pi\sum\limits_{i=1}^{N}Z_i^2=const.
\end{equation}
We assume non-trivial $\Xi>0$.

  Substitution $r=\exp(t)$ transforms (\ref{Fisher_1}) into  autonomous equation
can be easily integrated yielding
  \begin{equation} \label{N_fields_A}
  \frac{dY}{dt}= Y  -\frac{\Xi}{Y} +A\,,
  \end{equation}
 where $A$ is an integration constant.  Besides $A$, the result contains one more integration constant. Both are defined on account of (\ref{limits_XY}), in particular $A=r_g=2M$. The final result is
  \begin{equation} \label{N_fields_Fisher_form}
\left[g_{-}(Y)\right]^{{(1-\nu)}/{2}}\left[g_{+}(Y)\right]^{ {(1+\nu)}/{2}}=r\,,
  \end{equation}
 where $g_{\pm}(Y)=Y+M\pm\sqrt{M^2+\Xi}$,  $\nu=M/\sqrt{M^2+\Xi}$. Here $Y(r)$ varies from  $\sqrt{M^2+\Xi}-M$ to infinity as $r$ varies from zero to infinity. This determines implicitly $Y(r)\ge\sqrt{M^2+\Xi}-M$ as a function of $r>0$.

 The metric components are
   \begin{equation} \label{N_fields_Fisher_metric}
 e^{\alpha}=(g_-/g_+)^\nu, \quad e^{\beta}=g_+g_-/Y^2,
   \end{equation}		
  and SF as a function of $Y$ is
   \begin{equation} \label{JNW_SF}
   \phi_i(Y)=\frac{Z_i}{2\sqrt{M^2+\Xi}}\ln \left(\frac{g_{+}(Y)}{g_-(Y)}\right)  .
   \end{equation}
This is  the Fisher solution \cite{Fisher}, the only difference  is due to the presence of $N$ fields in (\ref{Ksi}) and (\ref{JNW_SF}).
    Transition to a new radial variable $Y$ leads to the Janis-Newman-Winicour \cite{Janis} (see also \cite{WymanVirbhadra}) representation  of the metric
   \begin{equation} \label{JNW_metric}
ds^2= \left(\frac{g_{-}}{g_+}\right)^\nu dt^2-\left(\frac{g_{+}}{g_-}\right)^\nu dY^2 - (g_{+})^{1+\nu}(g_{-})^{1-\nu}dO^2 .
\end{equation}
\begin{figure}[h]
	\vskip1mm
	\includegraphics[width=80mm]{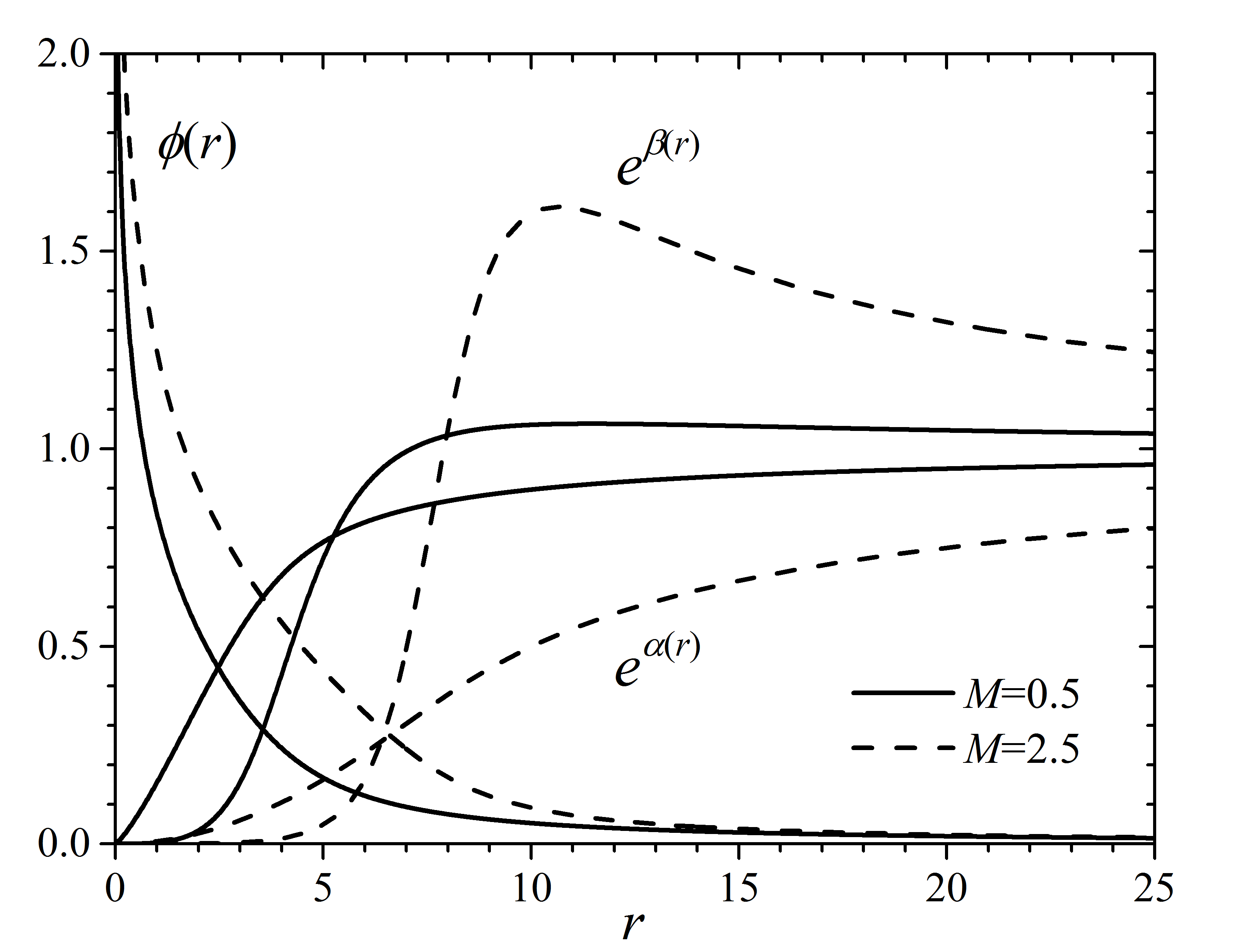}
	\vskip-3mm\caption{Metric and SF in case of the monomial potential (\ref{monomial0}) with $n=2$; the solutions are specified for $Q=0.5$ and different $M$. For $r\to 0$ it is  clearly seen $e^{\alpha}>>e^{\beta}$ in accordance with (\ref{asymptotics_alpha_beta}). }\label{Metrics}
\end{figure}
\begin{figure}[h]
	\vskip1mm
	\includegraphics[width=80mm]{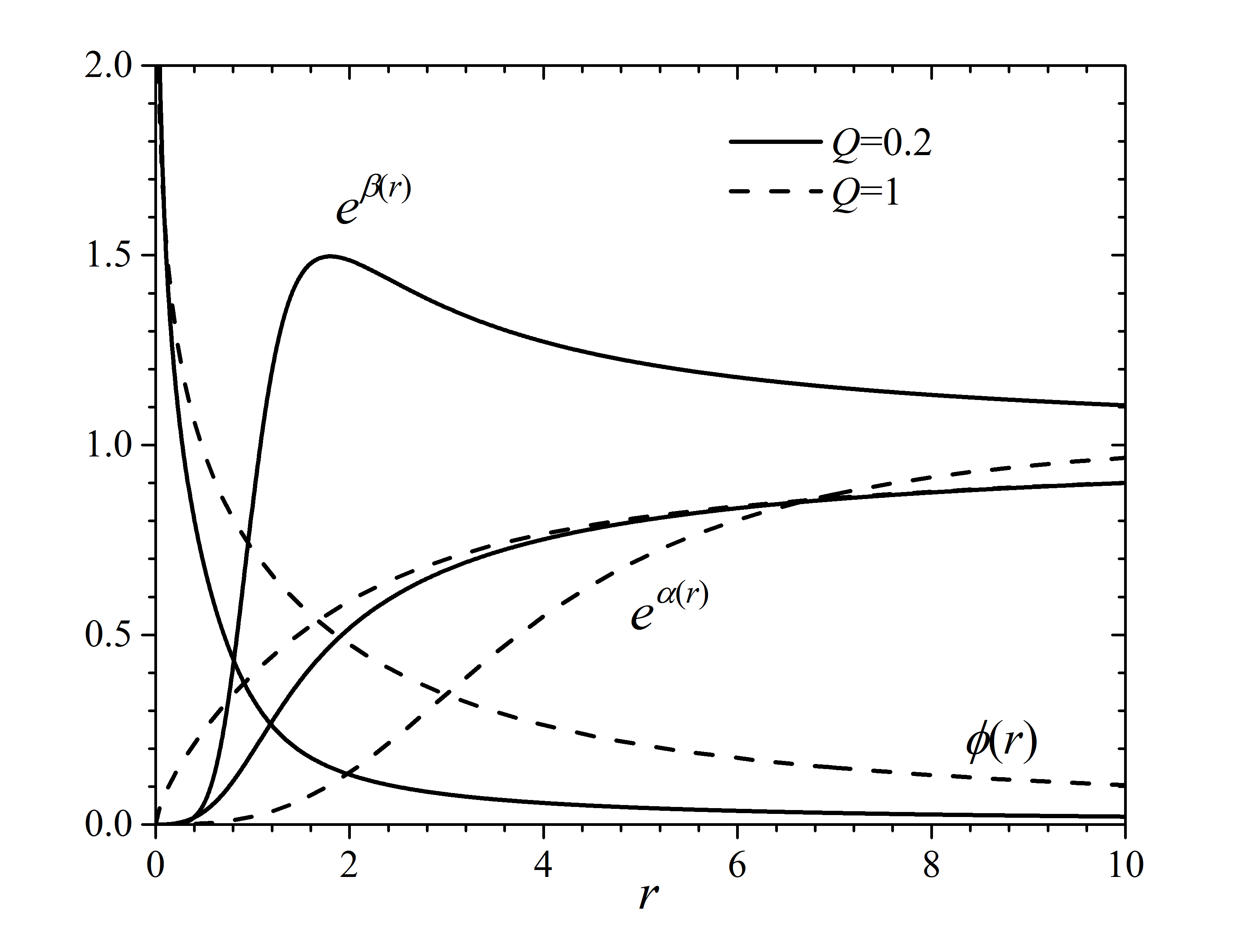}
	\vskip-3mm\caption{Metric and SF in case of the monomial potentials (\ref{monomial0}) with $n=3$; the solutions are specified for $M=1$ and different $Q$. }\label{Metrics-1}
\end{figure}

 \section{Numerical solutions: one field, monomial potential} \label{Section_asymptotic+numerical}
 \setlength\parindent{14pt}   Here we consider one SF with monomial potential
 \begin{equation}
 \label{monomial0}
 V(\phi)=\phi^{2n} ,
 \end{equation}
 where $n=2,3,...$.  The case of the linear massive scalar field  ($n=1$) has been considered in \cite{AsanovStashko3}. Obviously, assumptions  (\ref{positive potential_0}, \ref{positive potential_dash}) and  (\ref{estimate_of_V0}, \ref{estimate_of_V}) are fulfilled  with appropriately chosen $C_0, C'_0$.  Therefore all the results of {\it Lemmas}  1--4 and   {\it Theorem} 1  are valid for solutions  of equations (\ref{Ein_1-0} -- \ref{equation-phi}) with the potential (\ref{monomial0}).

 The asymptotic formulas for sufficiently large $r$  are obtained on account of conditions (\ref{infinity_limit_phi1}, \ref{infinity_limit_phi}, \ref{limits_XY}), assuming that the behavior of $\phi(r)$ for $r\to \infty$ must be the same as in the Minkowski space-time (see Appendix \ref{SF_in_flat_space}).
 For $n=3,4,...$  the asymptotic formulas can be derived using expansions in powers of $1/r$. The leading terms of solutions are as follows:
 \begin{equation}\label{asymptotics_inf_phi}
 \phi(r)=\frac{Q}{r}\left[1+\frac{r_g}{2r}+O\left(\frac{1}{r^3}\right)
 \right]\,,
 \end{equation}
 \begin{equation}
 \label{asymptotics_inf_metric-alpha}
 e^{\alpha}=\left(1-\frac{r_g}{r}\right)\left[1+O\left(\frac{1}{r^3}\right)\right] \,,
 \end{equation}
 \begin{equation}
 \label{asymptotics_metric-inf_beta}
 e^{ \beta}= \left(1-\frac{r_g}{r}\right)^{-1}\left[1-\frac{4\pi Q^2}{r^2}+O\left(\frac{1}{r^{3}}\right)\right]\,,
 \end{equation}
 where constants  $Q$ and $M$ fix the solution uniquely.
 For $n=2$ the asymptotic formula for SF involves logarithmic terms according to (\ref{phi_inf_4}). We note that in fact only the first terms of the asymptotic formulas are sufficient to obtain stable numerical results described below. Namely, one can use (\ref{flat_phi_more4}) and (\ref{phi_inf_4}) to obtain initial conditions for the numerical integration .
  of equations (\ref{Ein_A_2}, \ref{Ein_B_2}, \ref{equation-phi_2}, \ref{equation-phi_2a}),   starting with a sufficiently large value of $r$ towards the center  $r=0$.

\begin{figure}[h]
 	\vskip1mm
 	\includegraphics[width=80mm]{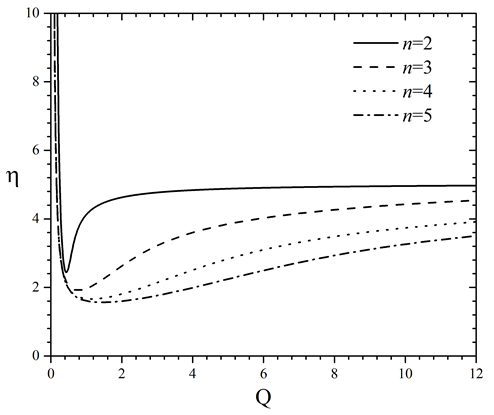}
 	\vskip-3mm\caption{Dependence $\eta(Q)$ for different  $n$ of  (\ref{monomial0}); $M=1$. }\label{Eta_of_Q_Mconst}
 \end{figure}

\begin{figure}[h]
 	\vskip1mm
 	\includegraphics[width=80mm] {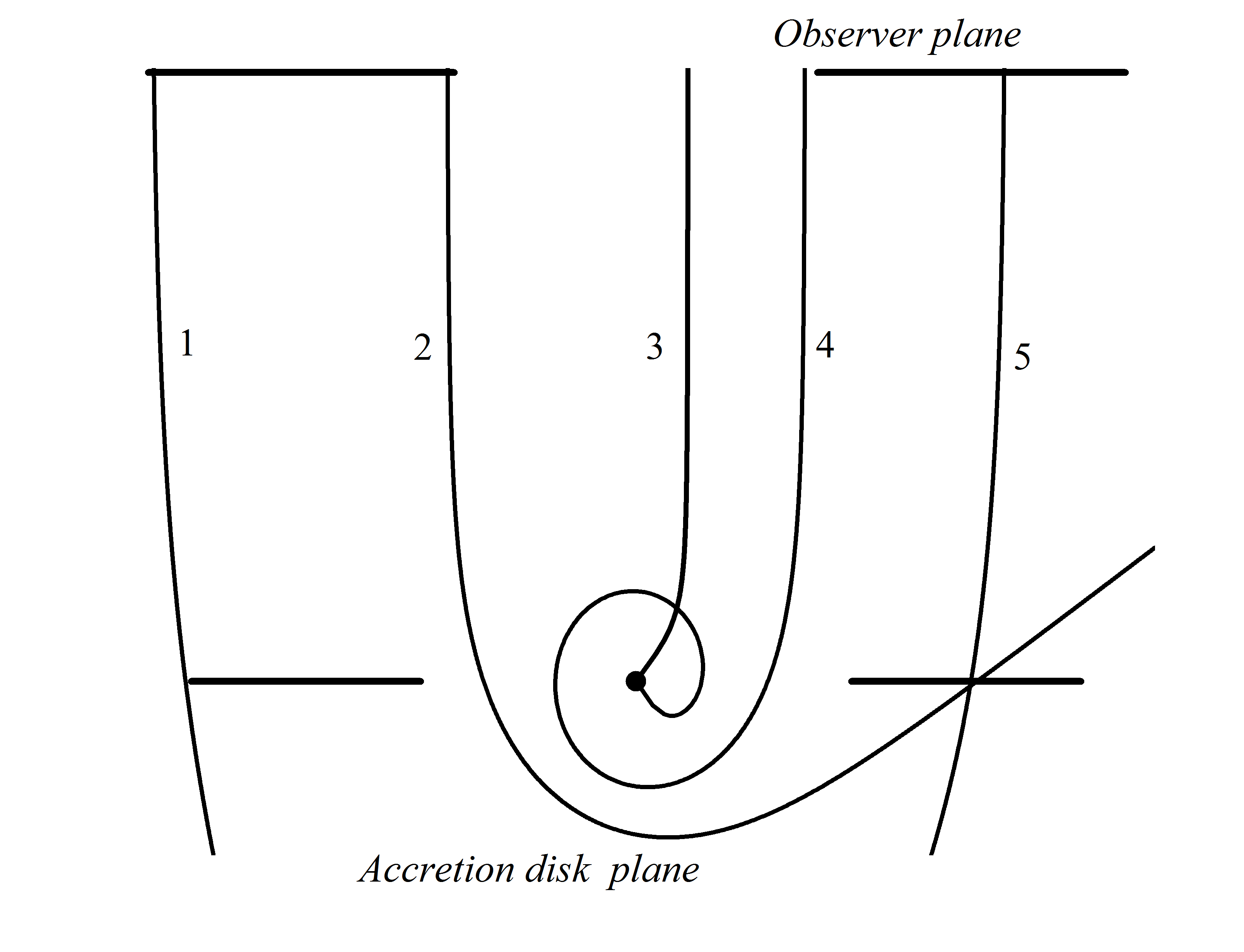}
 	\vskip-3mm\caption{Light rays incident from infinity in the field of the naked singularity; $\eta=4\pi \zeta_0^2>3$ ($n=3$, $Q=0.2$). The accretion disk is observed face-on; the empty section on the AD plane indicates an area in which  there are no SCOs forming the accretion disk. The fall of photons to the center with a sufficiently small $L$ is possible (trajectories 3,4). Trajectory 2 intersects the empty section of the AD plane and creates an image of a point on the back of AD.}\label{Ray-tracing_fall}
 \end{figure}

\begin{figure}[h]
 	\vskip1mm
 	\includegraphics[width=80mm] {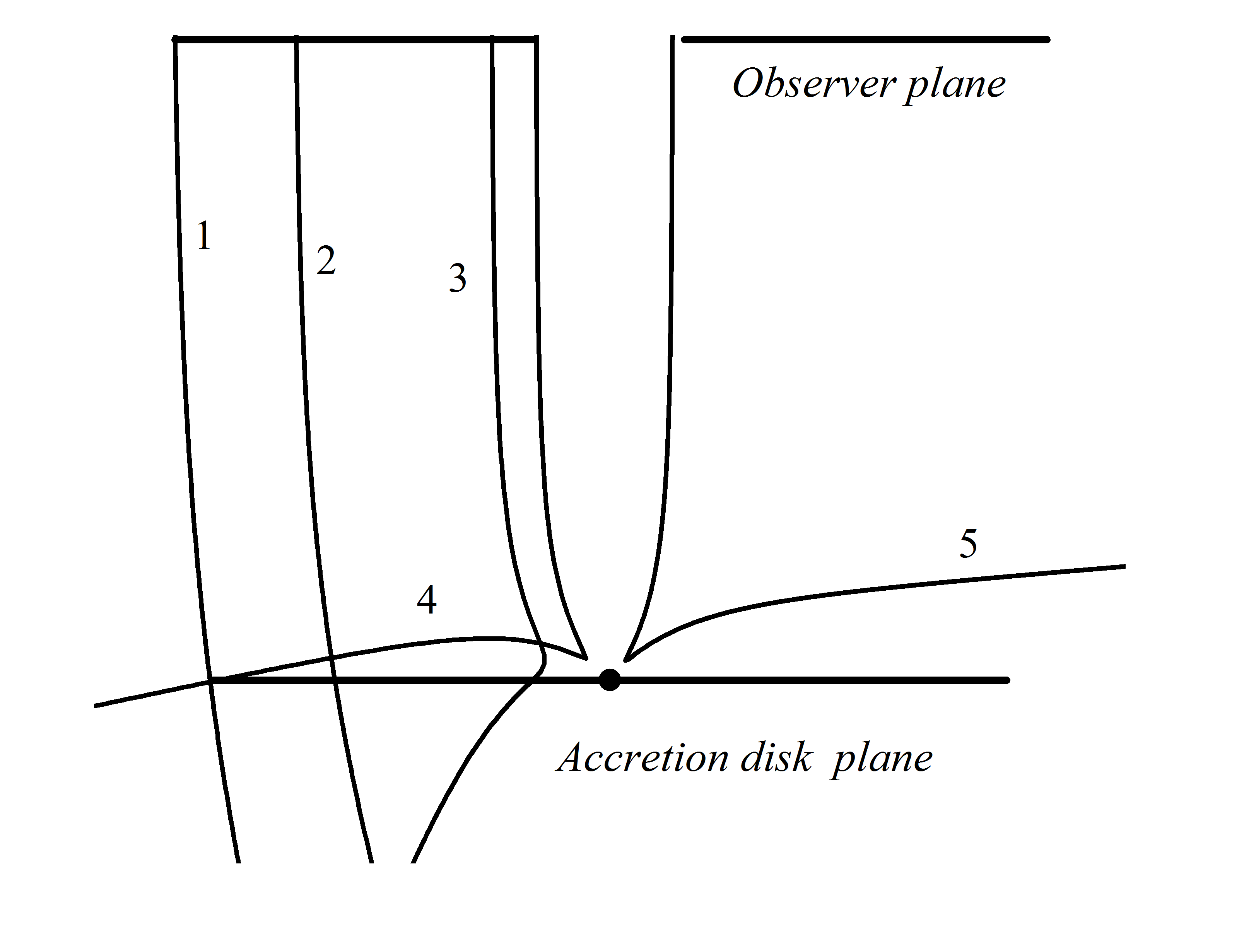}
 	\vskip-3mm\caption{Typical  photons trajectories in the field of the naked singularity for $\eta =4\pi \zeta_0^2<3$ ($n=3$, $Q=0.5$). Each point of the accretion disk has two images, with the exception of the area in the vicinity of the center; the photons from this region do not come to the observer plane. Trajectories 1 and 4 correspond to the different images of the disk boundary. In case of trajectories   4, 5 there is a strong deflection when the scattered  photons pass nearby center. Trajectory 5 does not cross the disk plane at all.}\label{Ray-tracing_no_fall}
 \end{figure}

 We checked the asymptotic behavior near the center ($r\to 0$), which  is described by the relations (\ref{asymptotics_phi}) for SF and (\ref{asymptotics_alpha_beta}) for the metric with $\eta=4\pi \zeta_0^2>0$.
 Figs. \ref{Metrics}, \ref{Metrics-1} illustrate  typical behavior of  the solutions. Fig. \ref{Eta_of_Q_Mconst} shows the relationship  $\eta(Q)$ of asymptotics at infinity and near the center. We see that both signs of $\eta-3$ are possible leading to different types of the photon trajectories (Figs. \ref{Ray-tracing_fall}, \ref{Ray-tracing_no_fall}).

 The numerical solutions have been used to study the geodesic structure around the configuration with potential (\ref{monomial0}) according to equations (\ref{geodesics_2}, \ref{1D particle}). The configuration parameters were $n=2, Q=0.2$ ($\eta>3$, Fig. \ref{Ray-tracing_fall}) and $n=3, Q=0.5$ ($\eta<3$, Fig. \ref{Ray-tracing_no_fall}); the figures show the  qualitative features  of the incident photon trajectories, which are typically used for imaging the configuration  by means of the inverse ray tracing.

 To illustrate, we considered a simple model of an accretion disk (AD) observed face-on, which is formed by the planar distribution of the test body stable circular orbits (SCO).  The SCO distribution have been studied  by means of the technique of our papers (see \cite{Stashko, Stashko4},  where the key point is the occurrence and disposition of extrema of $U_{eff}(r,L,1)$.
 
 In case of $\eta>3$,  we  have the Schwarzshild-like SCO distribution of AD \cite{Stashko4}: there is the inner region where the test body circular orbits do not exist at all, then there exists a ring of  unstable circular orbits with larger radii and then there is an outer region of SCO that extends to infinity. These regions are indicated by the empty section of the AD plane in Fig. \ref{Ray-tracing_fall}.

 For $\eta <3$, SCO densely fill the area near the center and there exist SCOs with arbitrarily small radii \cite{Stashko4} in the accretion disk plane, which is described by bold solid black line without gaps in Fig. \ref{Ray-tracing_no_fall}. Also in this case there may be discontinuous SCO distributions: in this case there is the region of SCO near the center, then there is a ring of unstable orbits and then there is an outer region of SCO that extends to infinity.

 A detailed description  of these results is beyond the scope of the present paper; however, we note that in case of the configurations with potential (\ref{monomial0}) fixed by parameters $M,\, Q$,  possible SCO distributions turn out to be qualitatively similar to the case of the linear massive scalar field (see \cite{Stashko4}).

 For $\eta<3$, there is a strong light deflection near the singularity and due to this effect the light from the inner orbits of AD near the center  avoids certain directions. The photons incident from infinity with a sufficiently small $L$ deviate strongly from the initial direction and do not reach the accretion disk that is located face-on; this means that the distant observer must see a dark spot in the center of the configuration. This does not mean that the innermost SCO will be invisible for {\textit all} possible observers since the effect depends on the  direction of the line of sight with respect to the AD plane.

 In case of $\eta>3$, the qualitative picture is as described in Section \ref{Section_Test_particle_motion}; in particular, the photons from infinity with sufficiently small $L$  fall to the singularity.

%\newpage

\end{document}